\documentclass[preprint2]{aastex}

\usepackage{amsmath}

\slugcomment{Submitted to Astrophysical Journal Letters}

\shorttitle{Potential Pluto Satellite Orbits}
\shortauthors{Porter \& Stern}

\begin{document}

\title{Orbits of Potential Pluto Satellites and Rings\\Between Charon and Hydra}

\author{S. B. Porter and S. A. Stern}
\affil{Southwest Research Institute, Boulder, CO 80302}
\email{porter@boulder.swri.edu}

\begin{abstract}
Pluto and its five known satellites form a complex dynamic system.
Here we explore where additional satellites could exist exterior to Charon (the innermost moon) but interior of Hydra (the outermost).
We also provide dynamical constraints for the masses of the known satellites.
We show that there are significant stable regions interior of Styx and between Nix and Kerberos.
In addition, we show that coorbitals of the known small satellites are stable, even at high inclinations,
and discuss mass constraints on undiscovered satellites in such orbits.
\\\\
\end{abstract}

\keywords{
    Kuiper belt objects: individual(Pluto)
    ---
    planets and satellites: individual(Pluto)
    ---
    planets and satellites: dynamical evolution and stability
    ---
    celestial mechanics
}

\section{Introduction}

The discovery of Pluto's major satellite Charon by \citet{cristy78} made Pluto-Charon the first known near-equal mass ratio \citep[8.2:1,][]{bro15} in the solar system,
and Charon remains the largest known satellite in the Kuiper Belt.
Pluto and Charon orbit their common barycenter in near-circular orbits with a period of 
6.3872 days and an average distance of 19,596 km \citep{bro15}.
\citet{stern91,stern94} searched for additional satellites from the ground and with the Hubble Space Telescope (HST), 
and while they did not find any satellites, the later study did 
discover the Charon Instability Strip, which destabilizes any any small prograde satellites within the 2:1 mean motion resonance with Charon.
\citet{weaver06} finally discovered two small satellites, Nix and Hydra, in 2005 using HST.
This was then followed up by Kerberos in 2011 \citep{show11} and Styx in 2012 \citep{show12}.
The four small satellites are closely packed together and are strongly perturbed by both each other and Charon \citep{you12}.
In addition, \citet{wint10,wint13,wint14} have shown that small satellites can be stable between Pluto and Charon, even at high eccetricities and inclinations.

NASA's \textit{New Horizons} spacecraft will fly through the Pluto system in July 2015, passing within 12,500 km of Pluto 
and 1,800 km of the Pluto-Charon L$_3$ point \citep{guo2008}.
Because of this close passage, the spacecraft will perform an intensive search for new satellites and rings of Pluto to make sure that 
there are no objects currently producing dust particles which could damage the spacecraft.
Though the likelihood of a new satellite being hazardous to the spacecraft can be shown to be
very low, the search for new satellites on approach will be the most complete survey of 
small Pluto satellites possible, far exceeding what can currently be done with ground-based and Earth-orbit observatories.
Here, we show where it is dynamically possible to insert new satellites with a small non-zero mass in the Pluto system
and speculate which of islands of stability are likely to be populated with currently-unseen satellites or rings.

\section{Numerical Methods}

To simulate the Pluto system, we implemented an n-body integrator in C++ with a Python interface.
This integrator is based on the 12(10) Runge-Kutta-Nystr{\"o}m method in ACM Algorithm 670 \citep{brak89},
and our implementation is available online\footnotemark.\footnotetext{\url{https://github.com/ascendingnode/PyNBody}}
This method is optimized for systems with a few strongly-interacting objects and no non-gravitational nor tidal dissipation forces.
We verified this method against an independent high-precision Bulirsch-Stoer integration of the Pluto system for 1000 years,
and found agreement between the two integrations to within 10 meters (D. P. Hamilton 2014, private communication).
For this implementation, we set a maximum distance from the barycenter of $10^5$ km, i.e., 
the simulation would be stopped if any object passed beyond this distance.
In addition, we assigned effective collision radii for the known and new small satellites of 500 km, 
so that the simulation would fail if any two small bodies passed within 1000 km of each other. 
While in reality such an encounter is not close enough to be an impact, 
we believe it is close enough to call into question the long-term stability of those two objects
(the Hill radii of the known small satellites are all $<$20 km).

\subsection{Known Satellite Uncertainties}

\citet{bro15} is the best published estimate of the masses and orbits of the known Pluto satellites.
However, while they provide mass uncertainties, 
the mass and state vectors of the satellites are coupled and certain masses are only possible with certain states.
We explored this parameter space by creating a dense discretely-sampled probability distribution function (PDF). 
We constructed an approximate log-probability function for a given state using the measurement uncertainty of the 
HST astrometric points in \citet{buie06,buie13} and \citet{bro15}, enforcing a minimum lifetime of $10^5$ Charon periods, and 
rejecting solutions where any of the objects passed within 1000 km of each other or one of the known satellite's barycentric eccentricity exceeded 0.1.
Since the uncertainty in Charon's orbit is much smaller, 
we used the best-fit Pluto and Charon masses and states from \citet{bro15} for all the simulations.
We then fed this function into the ``emcee'' Affine Invariant Markov chain Monte Carlo (MCMC) Ensemble sampler \citep{formac13} to build the PDF.
The result was an ensemble of probability-weighted possible known-satellite mass/state systems from which we could draw.
Because we enforced a minimum lifetime, our median masses for the known satellites were slightly different than \citet{bro15}, 
but within both their error bars and the stability constraints of \citet{you12}.

\subsection{Simulations of New Satellites}

We simulated unknown/unseen objects in the Pluto system adding an object with a 
specified non-zero mass and randomized state vector less than a given inclination,
verifying that it does not make the residuals of the known satellites significantly worse ($<$5\% increase in $\chi^2$),
and making sure that it has a lifetime greater than $10^5$ Charon orbits.
We assumed the new satellites had a gravitational mass of $10^{-5}$ km$^3$ s$^{-2}$, about 10\% of the mass of our best fit for Styx.
To create a random state vector, we first generated two random unit vectors for barycentric position and velocity using the method in \citet{knop70} and made sure that their cross product 
(i.e. the direction of the satellite's angular momentum vector) was angled less than a given inclination from the Pluto-Charon orbit pole.
For the final simulations, we generated cases with $<$10$^\circ$ inclination (to fully explore coplanar satellites) 
and a general case with unrestricted inclination.
We then generated a uniformly random magnitude for the position vector between 20,000 km (just outside the orbit of Charon) and 100,000 km 
(beyond the orbit of Hydra).
We calculated the circular orbital velocity at this distance from the barycenter and then generated a random velocity between 0 and 1.5 
times that circular velocity.
We also made sure that the period of this orbit was less than seven times that of Charon, as orbits beyond that point are not 
significantly influenced by the known small satellites.
For each state vector that passed these tests, we next added to the known objects of the Pluto system
and checked that the resulting $\chi^2$ was smaller than 1.05 times the $\chi^2$ of the solution in \citet{bro15}.
Finally, objects which survived for the ten-year span of observations were integrated forward for $10^5$ Charon orbits to verify 
their medium-term stability against close encounters with the known satellites.
Any test objects that passed within 1000 km of a known satellite were deemed unstable and rejected.

\section{Results}

\begin{figure}
    \plotone{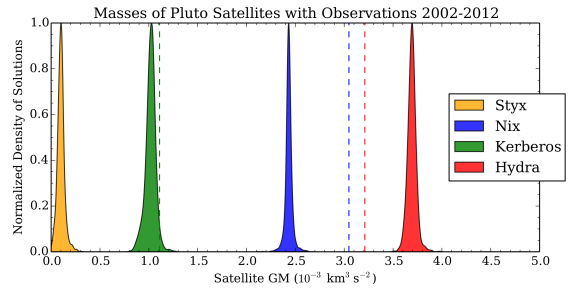}
    \caption{Derived mass uncertainties of the known Pluto satellites, 
        uncertainty from observations and constrained to solutions stable for $>$10$^5$ Charon orbits.
        Histograms represent our normalized mass distribution for each satellite,
        and the dashed lines indicate the masses from \citet{bro15}.
    \label{fig1}}
\end{figure}

\begin{figure}
    \plotone{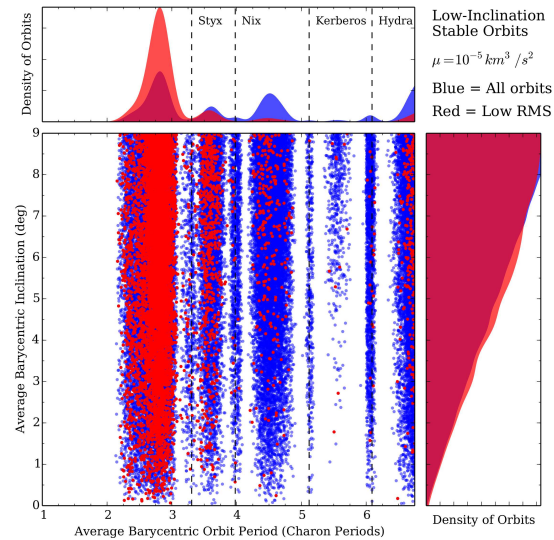}
    \caption{Distribution of 10$^5$ possible low-inclination orbits.
        The blue points show all the orbits that were stable for $>$10$^5$ Charon periods.
        The red points are the 10\% of the simulations with the lowest astrometric residuals, 
        and so are the most likely orbits for an additional low-inclination satellite.
        The near-coplanar islands interior of Styx and between Nix and Kerberos are the most stable regions for rings.
    \label{fig2}}
\end{figure}

\begin{figure}
    \plotone{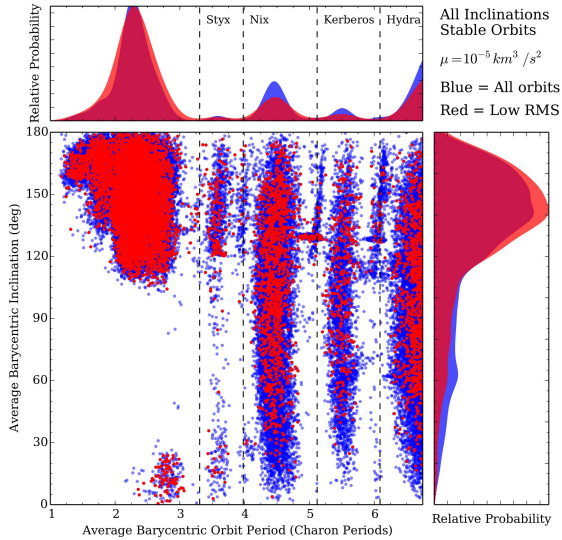}
    \caption{Distribution of possible high-inclination orbits.
        The blue points show all the orbits that were stable for $>$10$^5$ Charon periods.
        The red points are the 10\% of the simulations with the lowest astrometric residuals.
        Retrograde-inclined orbits appear more stable simply because they interact less with the known satellites;
        such satellites could only plausibly be formed through a very low-probability capture event.
    \label{fig3}}
\end{figure}

Our assumed mass distribution for the known satellites is shown in Figure \ref{fig1}, and compared to previous estimates in Table \ref{tab1}.
Figure \ref{fig2} shows the stable near-coplanar barycentric orbits, while Figure \ref{fig3} shows the orbits that were unrestricted in inclination.
In Figures \ref{fig2} and \ref{fig3}, the blue points are all the orbits, while the red points are the 10\% with the lowest astrometric residuals.

\subsection{Refined Known Satellite Masses and Implications}

We were able to obtain much tighter constraints on the masses of the known satellites than previous studies
by enforcing stability for $10^5$ Charon orbits (1747 years), rather than 200 years in \citet{bro15}.
Our largest divergence from \citet{bro15} was that by requiring Styx to be both stable and have a positive, non-zero mass,
we obtained a smaller mass for Nix (Nix is the largest perturber of Styx after Charon).
The smaller mass of Nix then required the mass of Hydra to grow, in order to place the same perturbations on Kerberos.
Our Kerberos mass is almost the same as \citet{bro15}, and supports the prediction in \citet{show15}
that Kerberos \citep[V=25.95,][]{show11} must have a significantly darker albedo than Nix and Hydra \citep[V=23.38 and V=22.93,][]{weaver06}.
The apparently dark surface of Kerberos is curious given that the small satellites can 
exchange surface material by sweeping up each other's low-velocity impact ejecta \citep{stern08,porter15}.
Our Hydra/Nix mass ratio ($\approx$1.5:1) is different to the mass ratio obtained by using the flux ratio and otherwise assuming equal 
properties ($\approx$1.8:1), but they are close enough that shape effects may be enough to account for the difference \citep{show15}.

\subsection{Possible Orbits of Near-Coplanar Satellites}

As shown in Figure \ref{fig2}, we found a great diversity of potential orbits for low-inclination unseen satellites 
which were both stable to the known satellites and did not adversely affect observed astrometric residuals.
The most stable region is just interior of Styx, centered just interior of the 3:1 mean motion resonance (MMR) with Charon.
Similar, slightly less stable regions are between Styx and Nix (around the 11:3 MMR) and Nix and Kerberos (around the 9:2 MMR).
As identified by \citet{you12}, the region between Kerberos and Hydra is very unstable.
Orbits beyond the 13:2 MMR were effectively unconstrained by the known satellites, 
while all prograde orbits interior of the 2:1 MMR were unstable, as shown in \citet{stern94}.

Regardless of period, we found that the number of stable orbits increases with inclination at a rate faster than cos(inclination).
This is due to lower-inclination orbits interacting more often and strongly with the known objects (which are all within 1$^\circ$ inclination), 
thus producing a more noticeable effect on the residuals.
Orbits interior of Styx are stable at the lowest inclinations without producing a noticeable effect on the residuals. 

We found that coorbitals of all the known satellites are also stable.
In the rotating frame of the known satellite they are coorbiting with, 
we found that most of these satellites are in low-eccentricity ``horseshoe orbits'' \citep{murray99} with large libration amplitudes,
though some (mainly Hydra coorbitals) were in ``tadpole orbits'' librating around the L$_4$ or L$_5$ Lagrange points.
Nix, Kerberos, and Hydra coorbitals with our assumed mass of $10^{-5}$ km$^3$ s$^{-2}$ had a negative effect on the residuals, so
any real coorbitals must be significantly lower mass than this.
However, Styx coorbitals with our assumed unknown satellite mass ($\approx$10\% Styx) had a much smaller effect on the residuals,
though this may be because we had much fewer astrometric points for Styx than the other satellites, so the problem is less well constrained.

\subsection{Possible Orbits of High Inclination and Retrograde Satellites}

As seen in Figure \ref{fig3}, we found stable high inclination orbits. 
At high (near-polar) inclinations, the region between Nix and Kerberos is most stable, followed by the region between Kerberos and Hydra.
Orbits interior of Nix at high inclinations are generally unstable.

There is also a huge stability plateau for retrograde satellites around the -7:3 MMR.
These orbits are stable to Charon and far enough interior of Styx that they do not affect the residuals.
In addition, there are curious Kerberos and Hydra coorbital structures at $\approx$140$^\circ$ inclination.

\section{Discussion}

We have shown that there are stable niches between Charon and Hydra at low inclination.
If there are any presently-undiscovered low-inclination 
Pluto satellites between Charon and Hydra, it is likely they formed in the same way as the known 
satellites, which are generally assumed to be remaining fragments from the Charon-forming impact \citep[and citations therein]{stern06,kenyon14}.
The most likely region for an unseen satellite is therefore in a low-inclination, low-eccentricity orbit with a period between 16 and 19 days.
Satellites up to 10\% of the mass of Styx can stably exist and not adversely affect the orbital residuals of the known satellites here. 
Low-inclination, low-eccentricity satellites with periods of around 24 and 29 days are also possible, but would need to be very low mass 
($<$1\% Styx) to avoid affecting the observational residuals.

This is in line with the detection limits from previous HST-based Pluto satellite searches, which placed strict limits on the brightness of objects 
between Nix and Hydra, but were much less sensitive interior of Styx due to the glare from Pluto \citep{steffl07}.
Since we were able to determine a dynamical mass for Styx of 10$^{-4}$ km$^3$ s$^{-2}$, our test mass of 10$^{-5}$ km$^3$ s$^{-2}$ is just small
enough to have plausibly passed detection previously.
Our test particles could have a noticeable perturbing effect on the real satellites for any orbit between Styx and Hydra,
suggesting that the dynamical limits to the size of any low-inclination satellites in this region are even more restrictive than the 
limits from HST-based searches.

NASA's \textit{New Horizons} spacecraft will perform a deep search for new satellites on approach to Pluto,
using the Long-Range Reconnaissance Imager (LORRI) with a sensitivity of greater than V=17 \citep{cheng08}
in order to assess the hazard to spacecraft from ejecta dust produced currently-undiscovered satellites.
The final hazard observations take place at a distance of less than 0.1 AU from the Pluto barycenter;
at this distance, LORRI is sensitive to objects down to $H_V>14$, or a few kilometers across.
\textit{New Horizons} could therefore discover satellites much smaller than our mass constraints.

We also found several classes of stable high-inclination and retrograde satellites. 
Some of these orbits, particularly those between Nix and Kerberos could exist at both low and high inclinations,
meaning a former low inclination satellite in this period range could be perturbed by an impact to higher inclination.
However, such a lucky impact is highly unlikely.

The large stable retrograde region interior of Styx could not easily by populated by remnants from the Charon-forming impact,
but could instead host captured satellites.
Binary-exchange capture \citep{funato04,agnor06} would happen when a binary KBO has very close encounter with the Pluto system, 
and one part of the binary is captured, while the other part is ejected from the system.
Such an event is very unlikely today, but would have been much more likely in the giant planet migration era, 
when Triton is presumed to have been captured around Neptune in exactly this fashion \citep{agnor06}.
Such a lucky stable capture is very unlikely, but if it happened, it would almost certainly be to a retrograde orbit with a period between 
13 and 19 days.

We did not perform any direct simulations of rings, as our simulations were only set up to simulate a single test particle.
However, Figure \ref{fig2} shows that there are several near-coplanar stable ``islands'' interior of Hydra,
especially interior of Styx and between Nix and Kerberos.
Observations by \citet{steffl07} using HST placed very tight upper limits on the brightness of any rings in the Nix/Hydra (I/F$<$10$^{-5}$),
and therefore it is not likely that the region between Nix and Kerberos is populated by a dense ring.
Thus the most likely region for a ring to exist around Pluto is interior of Styx, just interior of the Charon 3:1 MMR.

\section{Summary}

\begin{enumerate}
    \item We determined new constraints for the masses of the known satellites of Pluto, including a lower mass for Nix, which allowed for 
        Styx to have realistically-constrained mass.
    \item We found that additional low-inclination satellites are stable interior of Styx, between Styx and Nix, and between Nix and
        Kerberos. Coorbitals of the known satellites are also possible, but must be very low mass.
    \item We found that retrograde orbits between the 2:1 and 3:1 MMRs with Charon were quite stable, though very unlikely with current 
        formation models of the Pluto-Charon system.
\end{enumerate}

\acknowledgments

This work was supported by the NASA \textit{New Horizons} mission (contract number \#NAS02008).
Special thanks to Marc Buie for providing the computing resources for this project,
as well as Mark Showalter, Douglas Hamilton, and John Spencer for insightful discussions.


\begin{deluxetable}{lcccc}
    \tablecaption{Masses of Known Satellites (10$^{-3}$ km$^3$ s$^{-2}$)\label{tab1}}
    \tablewidth{0pt}
    \tablehead{ 
        \colhead{Satellite} & \colhead{Tholen et al.} & \colhead{Youdin et al.} & \colhead{Brozovi{\'c} et al.} & \colhead{This}\\ 
        \colhead{} & \colhead{(2008)} & \colhead{(2012)} & \colhead{(2015)} & \colhead{Work\tablenotemark{a}}
    }
    \startdata
    Styx         & - & - & $\le1.0$ & $0.10\pm0.03$ \\
    Nix          & $39\pm34$ & $\le3.3$ & $3.0\pm2.7$ & $2.43\pm0.02$ \\
    Kerberos     & - & - & $1.1\pm0.6$ & $1.02\pm0.05$ \\
    Hydra        & $21\pm42$ & $\le6.0$ & $3.2\pm2.8$ & $3.69\pm0.04$ \\
    \enddata
    \tablenotetext{a}{1-$\sigma$ uncertainty of solutions stable for $>$10$^5$ Charon orbits}
\end{deluxetable}

\end{document}